%% file: RJwrapper.tex
\documentclass{article}

\usepackage{arxiv}

\usepackage[utf8]{inputenc} 
\usepackage[T1]{fontenc}    
\usepackage{hyperref}       
\usepackage{url}            
\usepackage{booktabs}       
\usepackage{amsfonts}       
\usepackage{nicefrac}       
\usepackage{microtype}      
\usepackage{lipsum}		
\usepackage{graphicx}
\usepackage{natbib}
\usepackage{doi}
\usepackage{amsmath,amssymb,array}
\usepackage{amsthm}
\usepackage{cleveref}

\newtheorem{remark}{Remark}
\newtheorem*{Example*}{Example}

\usepackage{enumitem}

\usepackage{booktabs}

\newtheorem{theorem}{Theorem}

\graphicspath{{figures/}} 
\title{SPLINE BASED METHODS FOR FUNCTIONAL DATA ON MULTIVARIATE  DOMAINS }

\date{} 					

\author{Rani Basna \thanks{All authors contributed equally} \\
        Department of Statistics, Lund University, Sweden,\\
	\texttt{rani.basna@stat.lu.se} \\
	\And
	Hiba Nassar\\
	Cognitive Systems, Department of Applied Mathematics and Computer Science,\\
        Technical University of Denmark, Denmark\\
	\texttt{hibna@dtu.dk} \\
	 \AND
	Krzysztof Podg\'orski \\
	Department of Statistics, Lund University, Sweden \\
	 \texttt{ Krzysztof.Podgorski@stat.lu.se} \\
}

\begin{document}
\maketitle
\begin{abstract}
Functional data analysis is typically performed in two steps: first, functionally representing discrete observations, and then applying functional methods to the so-represented data. 
The initial choice of a functional representation may have a significant impact on the second phase of the analysis, as shown in recent research, where data-driven spline bases outperformed the predefined rigid choice of functional representation. 
The method chooses an initial functional basis by an efficient placement of the knots using a simple machine-learning algorithm. 
The approach does not apply directly when the data are defined on domains of a higher dimension than one such as, for example, images. The reason is that in higher dimensions the convenient and numerically efficient spline bases are obtained as tensor bases from 1D spline bases that require knots that are located on a lattice. This does not allow for a flexible knot placement that was fundamental for the 1D approach. 
The goal of this research is to propose two modified approaches that circumvent the problem by coding the irregular knot selection into their densities and utilizing these densities through the topology of the spaces of splines. This allows for regular grids for the knots and thus facilitates using the spline tensor bases.   
It is tested on 1D data showing that its performance is comparable to or better than the previous methods. 

\end{abstract}


\input{RJtemplate}

\end{document}

%% file: RJtemplate.tex
\section{Introduction}
\label{intro}

\begin{figure}[t!]
\begin{center}
\includegraphics[width=0.43\linewidth]{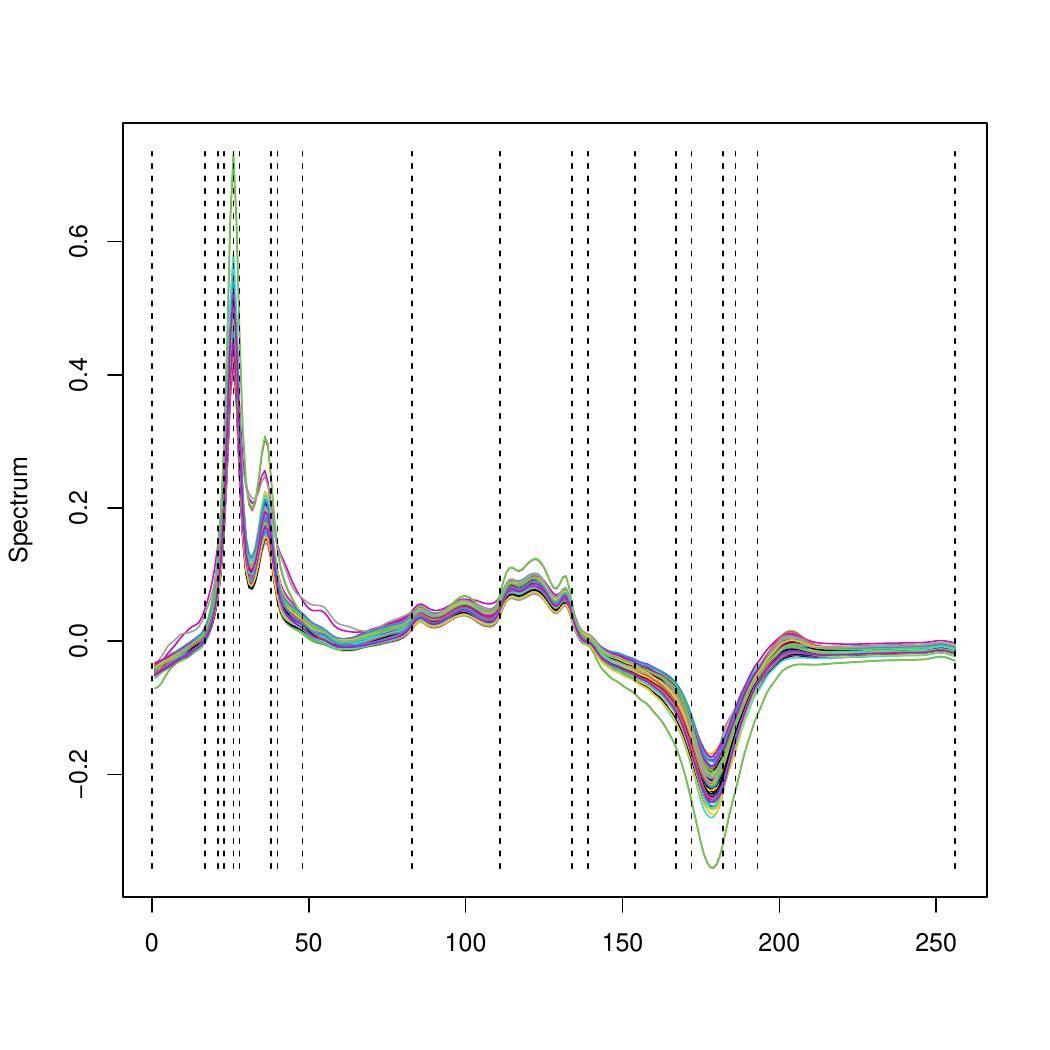} 
\includegraphics[width=0.43\linewidth]{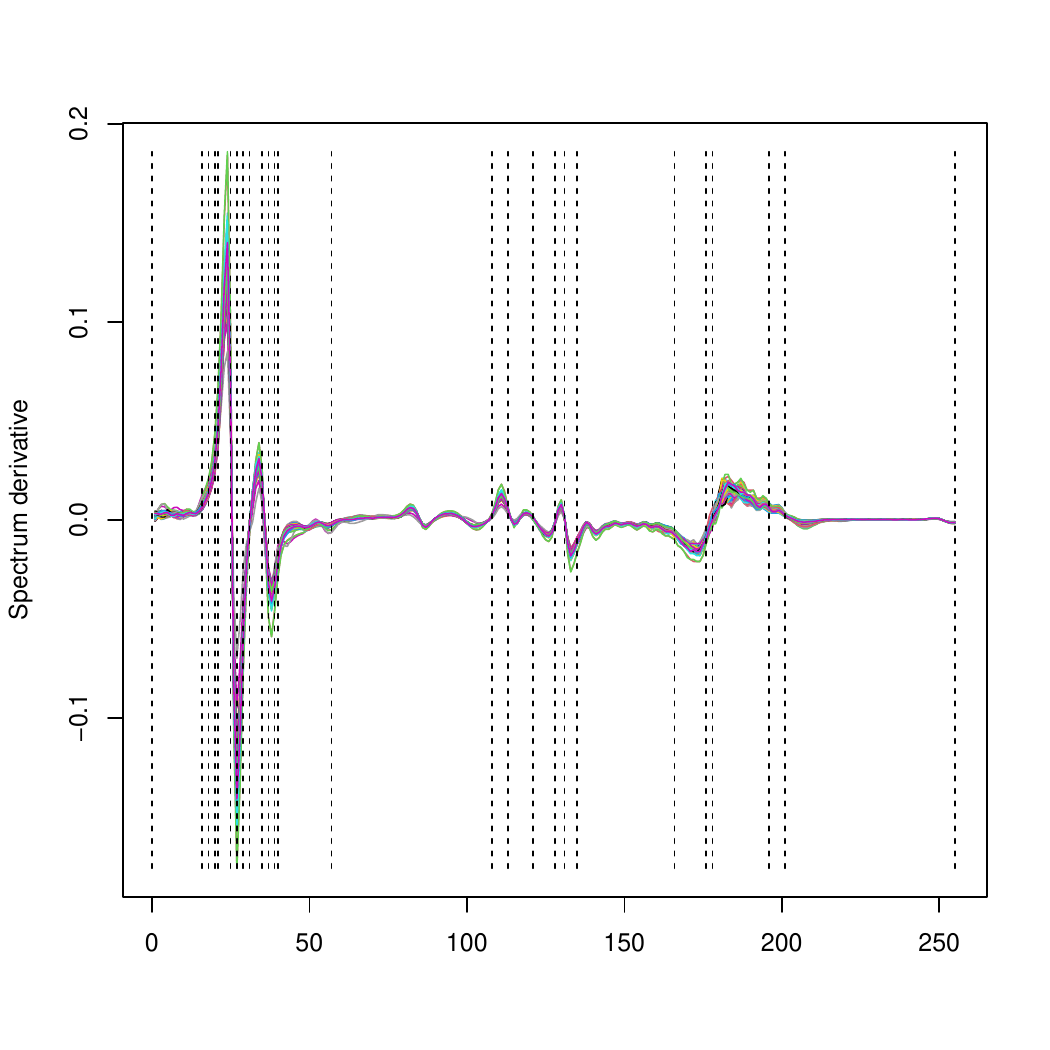} \\
\includegraphics[width=0.43\linewidth]{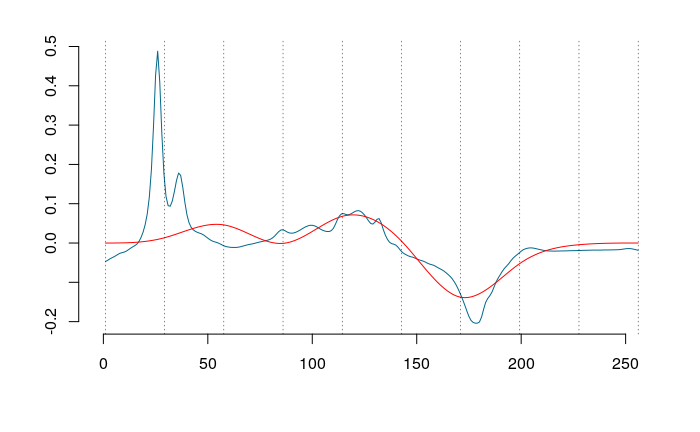} 
\includegraphics[width=0.43\linewidth]{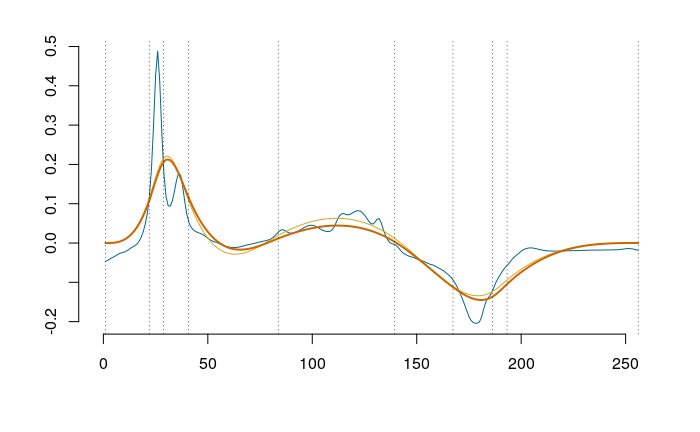} 
\end{center}
\caption{1D data and the knot selection. {\it (Top)}:  1D case, the wine data {\it (left)} and their derivatives {\it (right)}, the knots are marked by vertical dashed lines. {\it (Bottom)}:  Projection to the first four principal component for equally spaced {\it (left)} and data-driven, optimally spaced knots {\it (right)}.}\label{fig:1Dvs2D}
\end{figure}

The data-driven knot selection algorithm (DDK),  see \cite{basna2022data}, identifies regions in the common domain of functional data that feature high curvatures and other significant systematic variability. 
By strategically placing knots in these regions, the algorithm effectively captures the systematic variation and is available as a package on the GitHub repository \href{https://github.com/ranibasna/ddk}{\cite{BasnaR2021}}. 
In the one-dimensional domain case, the placement of knots can be directly utilized in the choice of the splines spanned over selected knots and it has been proved that this improves the efficiency of the functional representation of the data and the ensuing statistical analysis. 
The reason behind the improvement is dimension reduction in the initial representation of the data achieved by a `clever' choice of the knots.
Figure~\ref{fig:1Dvs2D}~{\it (top)} presents the knot selection for the wine spectra {\it (left)} and their derivatives {\it (right)}, see \cite{Wines}. 
Once the knots are provided, one can build the splinet i.e. a convenient orthonormal spline basis spanned over these knots, \cite{podgrski2021}. How the choice of the knots can improve the analysis of the data is shown in Figure~\ref{fig:1Dvs2D}~{\it (bottom)}, where we see the projection of the data to the first four principal components for equally spaced knots {\it (left)} and for the optimized data-driven knot selection {\it (right)}.

In 2D, one can utilize tensor product spline bases only, since the knots reside on the regular lattice knots. On the other hand, if one chooses `knots' to optimize the mean square error (MSE), the selected `knots' are irregular and they cannot be used directly to build convenient spline bases. 
In this note, we discuss how an irregular `knot' selection can be utilized to modify the approach that allows the use of the tensor bases but at the same time accounts for the data-driven `knot' selection. 
In fact, the bases will be constructed on equally spaced knots which have an additional computational advantage.
The approach is through utilizing the obtained irregular `knot' distribution to change the topology of spline spaces. 
The two methods of doing this will be considered: the first one is based on changing the topology of the state space (the space of values of splines) and the second one is based on transforming the domain space of splines. 

The note is organized as follows. 
In the next section, we briefly recap basic facts about the orthonormal basis of spline called {\em splinets}, which were recently introduced in \cite{LIU2022}. 
There, we also define two dimensional orthonormal tensor splinets. 
This is followed by a section on the proposed method of the knot selection in 2D. The two methods considered in this note are described in the two subsequent sections. First, the state-space transformation method is discussed in the 1D case and then explain the analogous approach to the 2D case. 
A similar scheme is followed in the section on the domain transformation method.
Next, we introduce the concept of classification for functional data in a $d$-dimensional domain. The note concludes with a description of empirical studies designed to test these methods.

 \section{Tensor splinets}
 \label{sec:tensor}
\begin{figure}[t!]
\begin{center}
\includegraphics[width=0.45\textwidth, height= 6cm]{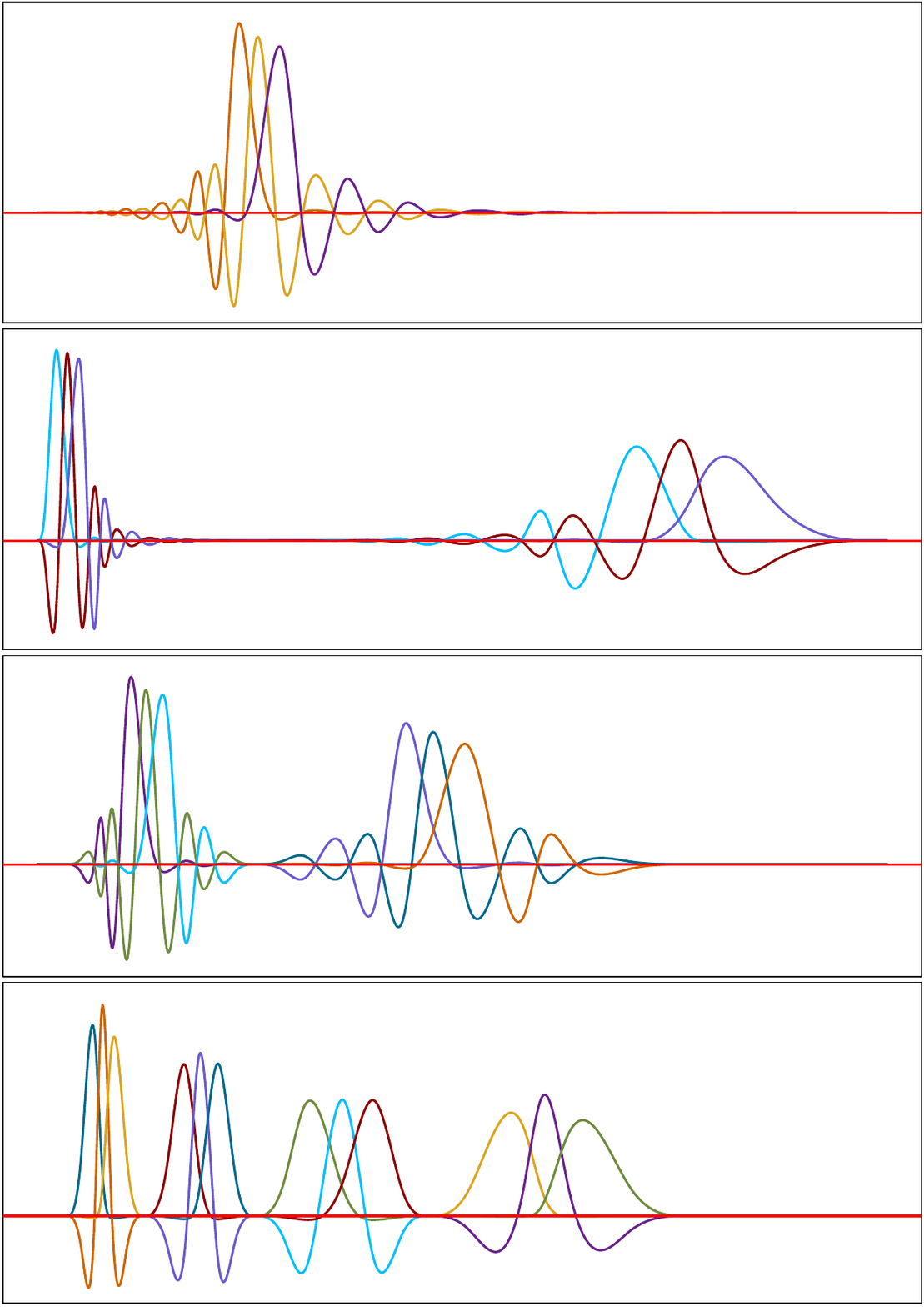}
\includegraphics[width=0.45\textwidth, height= 6cm]{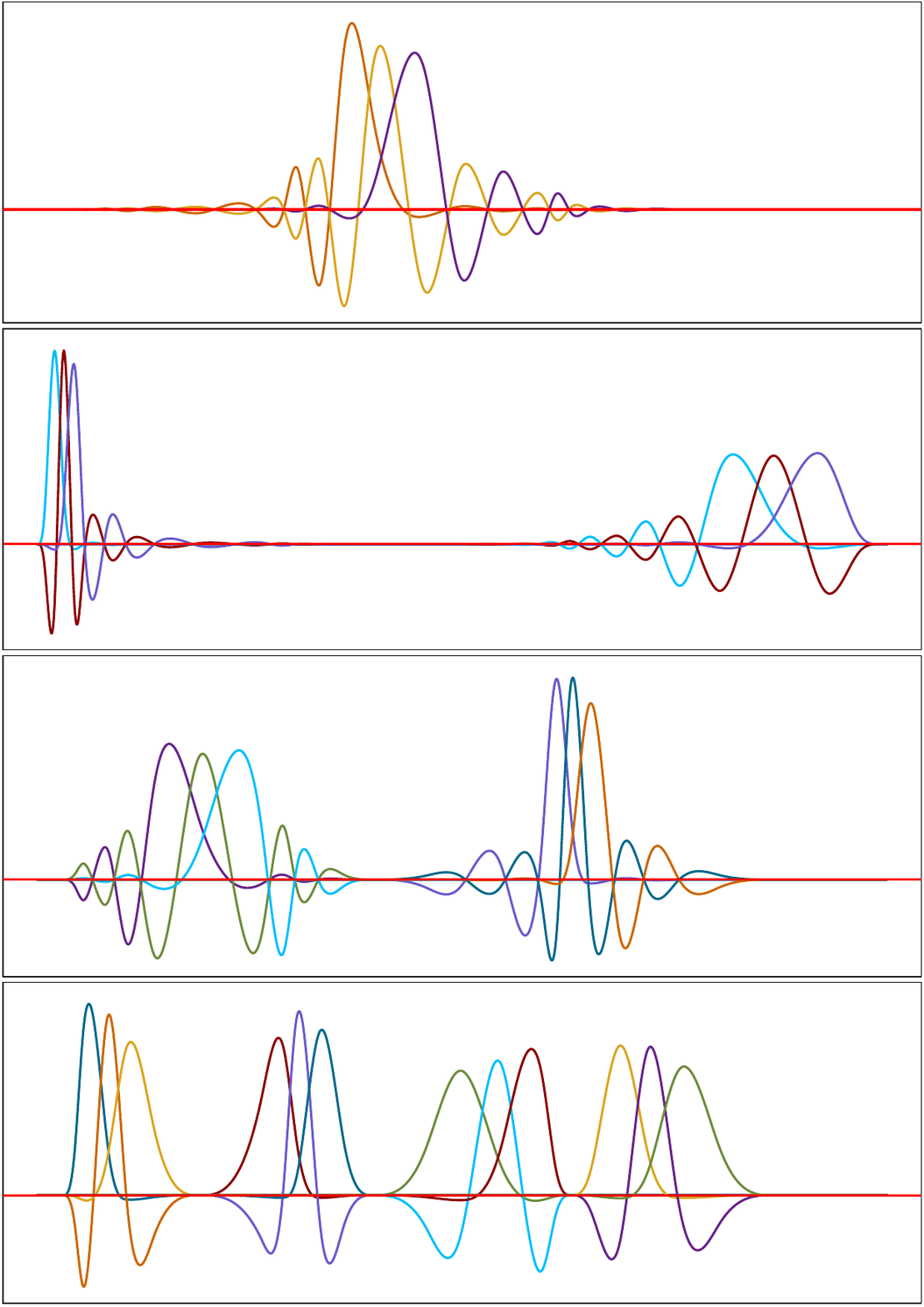}
\end{center}
\label{fig:Splinetdyadic}
\caption{ Examples of splinets of the third order in the graphical representation of the dyadic pyramid.
In the figure, we see that the placement of knots changes the ability to model local variability in the data. }
\end{figure}

 The splinets are efficient orthonormal bases of splines discussed in full detail in \cite{LIU2022} and implemented in the $R$-package {\it Splinets}, \cite{LiuNP}.
 For our purposes, it is enough to know that a splinet is obtained by an efficient dyadic algorithm from $B$-splines spanned on a given set of knots. Two examples are shown in Figure~\ref{fig:Splinetdyadic}, where the effect of different knot distributions is seen. 

 In the package, the focus is on one-dimensional domains. 
 However, the tensor product bases are simple enough so that the package can also be used for the analysis of functional data defined on an arbitrary $d$-dimensional space. 

Let us consider $d$ knot selections $a_j=\xi_{0,j} < \dots <\xi_{n_j+1,j}=b_j$, $j=1,\dots, d$ that constitute a lattice partition on the cube $\bigotimes_{j=1}^d[a_j,b_j]$.
We refer to these selections of knots as $\boldsymbol \xi_j$, $j=1,\dots, d$.
Further, let us consider $d$ sets $\mathcal{B}_{\boldsymbol \xi _j}^{k_j}$ of $B$-splines built on these knot selections (of possibly different smoothness $k_j$, $j=1,\dots,d$) denoted by 
$$
\mathcal{B}_{\boldsymbol \xi _j}^{k_j}=\left\{[a_j,b_j]\ni x\mapsto B_{\boldsymbol \xi_{j}}^{i,k_j}(x); i=1,\dots,n_j-k_j+1 \right\},
$$
$j=1,\dots, d$. 
Here, the index $i$ refers to the $i$th  $B$-spline in the linear ordering of the $B$-spline basis.
The linear functional spaces spanned on these collections of $B$-splines are denoted by $\mathcal H^{k_j}_{\boldsymbol \xi_j}$, $j=1,\dots,d$.
Then the tensor product $B$-spline basis $\bigotimes_{j=1}^d\mathcal{B}_{\boldsymbol \xi _j}^{k_j}$ is defined for $\mathbf x=(x_1,\dots,x_d)\in\bigotimes_{j=1}^d[a_j,b_j]$ through
\begin{align}
\left(\bigotimes_{j=1}^dB_{\boldsymbol \xi_j}^{i_j,k_j}\right)(\mathbf x) =  \prod_{j=1}^d B_{\boldsymbol \xi_j}^{i_j,k_j}(x_j).
\end{align}
Let $\bigotimes_{j=1}^d\mathcal H^{k_j}_{\boldsymbol \xi_j}$ be the $\prod_{j=1}^d(n_j-k_j+1)$-dimensional linear functional space on $\bigotimes_{j=1}^d[a_j,b_j]$ spanned by the tensor product basis. 

\begin{theorem}
\label{th:project}
Let $f$ be a square-integrable function on $\bigotimes_{j=1}^d[a_j,b_j]$ and the splinets $\{\mathcal{OB}_{\boldsymbol \xi _j}^{k_j}\}$, $j=1,\dots, d$ are orthogonormal splines bases obtained by dyadic orthogonalization the corresponding $B$-spline bases, as introduced in \cite{LIU2022}.  Then  
$\bigotimes_{j=1}^d\mathcal{OB}_{\boldsymbol \xi _j}^{k_j}$ 
also constitute an orthonormal basis in $\bigotimes_{j=1}^d\mathcal H^{k_j}_{\boldsymbol \xi_j}$ and 
the projection $\hat f$ of $f$ to this space is given by
\begin{align*}
\hat f&=\sum_{i_1=1}^{n_1-k_1+1}\dots \sum_{i_d=1}^{n_d-k_d+1} \alpha_{i_1\dots i_d}\cdot \bigotimes_{j=1}^d{OB}_{\boldsymbol \xi _j}^{i_j,k_j},\\
\alpha_{i_1\dots i_d}&=\int_{a_1}^{b_1} 
OB_{\boldsymbol \xi_1}^{i_1,k_1}(x_1)
\dots \int_{a_d}^{b_d} OB_{\boldsymbol \xi_d}^{i_d,k_d}(x_d) f(\mathbf x)~dx_d\dots~dx_1.
\end{align*}

Moreover, for each function $f$ satisfying zero boundary conditions of the $k_j$th order along the $j$th coordinate and each $\epsilon > 0$ there exist knot selections $\boldsymbol \xi_j$, $j=1,\dots, d$ such that 
$$
\left(\int_{\bigotimes_{j=1}^d[a_j,b_j]} \left|f(\mathbf x)-\hat f(\mathbf x) \right|^2~d\mathbf x \right)^{1/2}\le \epsilon.
$$
\end{theorem}

\begin{figure}[t!]
\begin{center}
\includegraphics[width=0.24\linewidth]{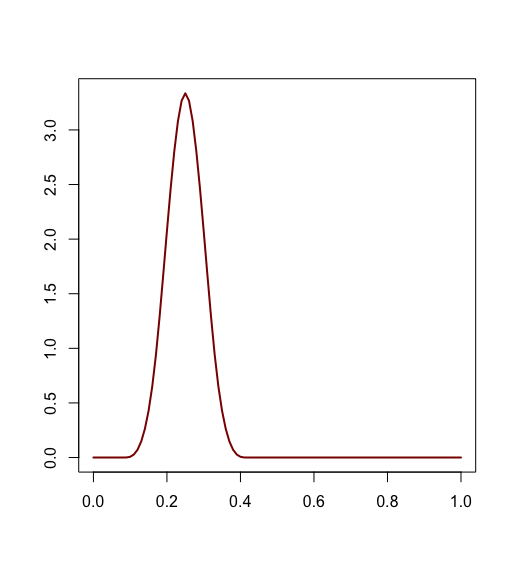}
\includegraphics[width=0.24\linewidth]{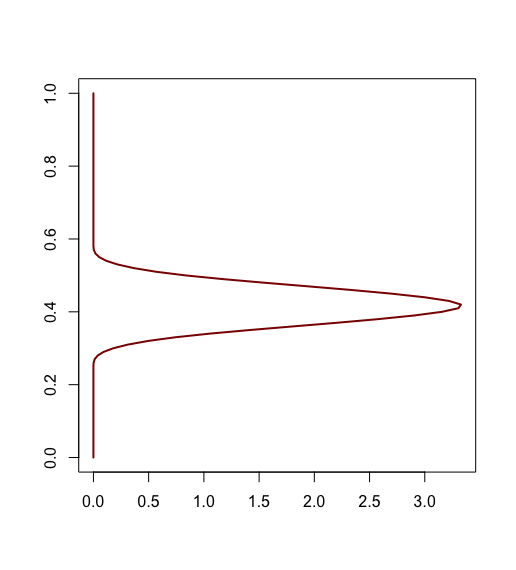}
\includegraphics[width=0.24\linewidth]{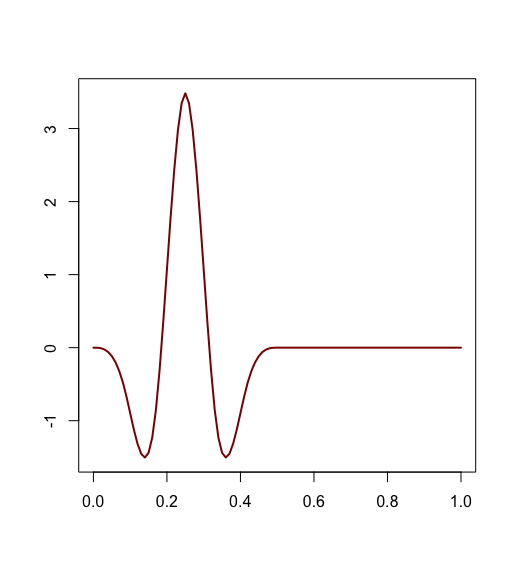}
\includegraphics[width=0.24\linewidth]{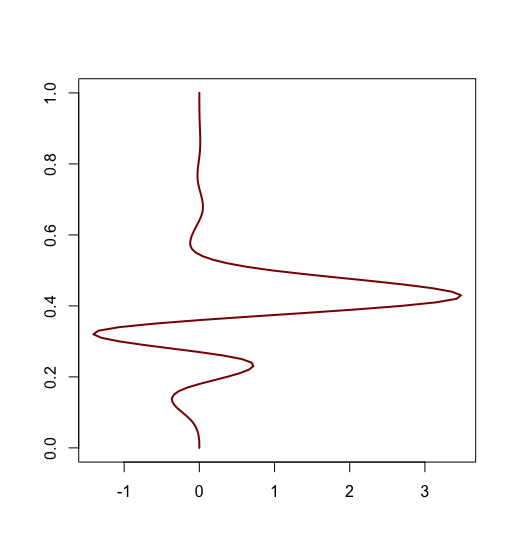}\\
\includegraphics[width=0.39\linewidth, height= 3.8cm]{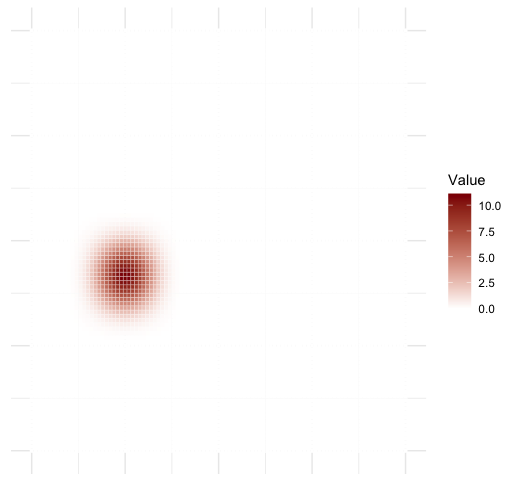}
\includegraphics[width=0.39\linewidth, height= 3.8cm]{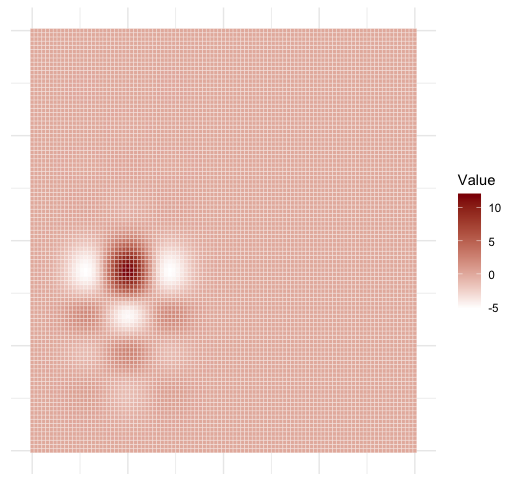}
\end{center}
\caption{
{\it Top:} Two $B$-splines {\it (left)} and corresponding $OB$-splines {\it (right)}. {\it Bottom:} The corresponding tensor spline $B_{\boldsymbol \xi_1}^{i_1,3}\otimes B_{\boldsymbol \xi_2}^{i_2,3}$
{\it (left)} and the orthogonal tensor spline $OB_{\boldsymbol \xi_1}^{i_1,3}\otimes OB_{\boldsymbol \xi_2}^{i_2,3}$ {\it (right)}. }\label{fig:Splinet}
\end{figure}

In Figure~\ref{fig:Splinet}, we present abscissa and ordinate 1D-basis elements at the top and an element of bivariate tensor $B$-spline $B_{\boldsymbol \xi_1}^{i_1,3}\otimes B_{\boldsymbol \xi_2}^{i_2,3}$ {\it (left)} and the corresponding $OB$-spline  $OB_{\boldsymbol \xi_1}^{i_1,3}\otimes OB_{\boldsymbol \xi_2}^{i_2,3}$ {(\it right)} at the bottom. 
\section{`Knot' selection in 2D}

\begin{figure}[t!]
\begin{center}
\includegraphics[width=0.32\linewidth]{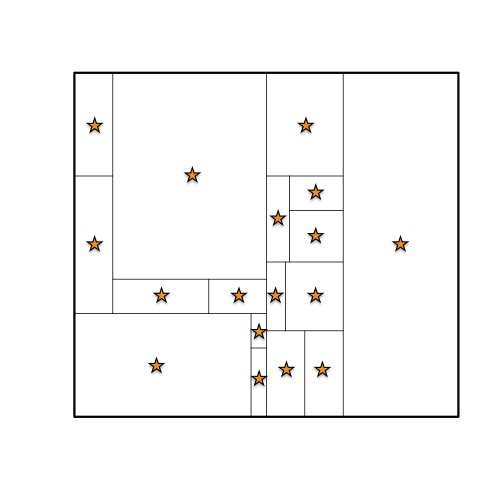}
\includegraphics[width=0.32\linewidth]{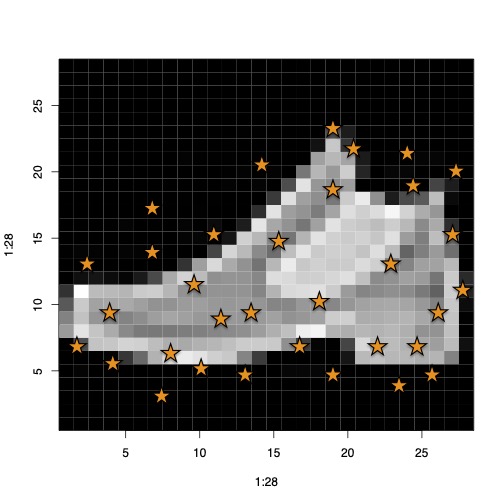} 
\includegraphics[width=0.32\linewidth]{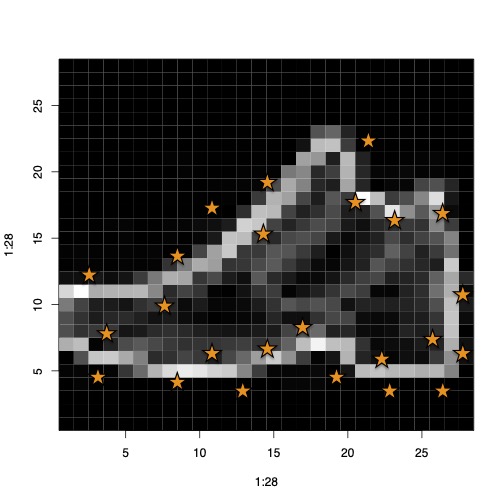} 
\end{center}
\caption{
{\it (Left)}: Illustration of how the `knots' in 2D are obtained by taking the midpoints of 2D rectangular grid cells obtained by the binary regression tree MSE fit to the functional data. {\it (Middle)} The original image and the selected `knots', {\it (Right)} The gradient image with `knots'. }\label{fig:2D}
\end{figure}

The algorithm that has been used for the 1D knot selection is applicable to functional data that are defined on the 2D plane. 
Indeed, the method is equivalent to building a regression binary tree where the so-called split points (the nods at the tree) are found based on the best mean square error (MSE) approximation of a regression function by functions that are constant over a 2D grid, see \cite{HastieTF9}. 
The centers of rectangles of the 2D grid define `knots' as shown in Figure~\ref{fig:2D}~{\it (left)}.
The `knots' for the MNIST fashion images and their gradient images are presented in Figure~\ref{fig:2D} {\it (middle, right)}, see \cite{MnistFashion} for more details on this data set.

We are using quotation marks when referring to obtained points since they are not convenient to build direct spline approximation similarly as in the one-dimensional case. 
The reason for this is that there are no convenient spline bases spanned on irregularly distributed `knots' in such domains. 
One could consider triangulation, but defining splines on a triangular grid is complicated, and the resulting spline bases are difficult to orthogonalize. 
The easiest to use and most popular 2D space of splines are based on the tensor product bases, see, for example, \cite{EilersM}. 
However, the tensor bases are defined over lattice grids, and therefore the 2D grids shown in Figure~\ref{fig:2D} {\it (Bottom)} cannot be used directly to construct such bases. 

However, from such grid centers one can estimate the intensities representing their concentrations on the plane or, in general, in $d$-dimensional domain. The methods of obtaining such intensities are not discussed in this work, but many existing methodologies can be implemented.  From now on it will be assumed that for data defined on a cube so that in a general $d$-dimensional setup we assume that for the functional data $f_i:\bigotimes_{j=1}^d[a_j,b_j]\mapsto \mathbb R$, $i=1,\dots,n$ through a certain algorithm the intensity of the grid center $g: \bigotimes_{j=1}^d[a_j,b_j]\mapsto [0,\infty]$, such that
$$
\int_{a_1}^{b_1}\dots \int_{a_d}^{b_d} g(x_1,\dots, x_d)=1
$$
has been estimated from the data.
In Sections~\ref{sec:state} and \ref{sec:domain}, we discuss how this intensity is used to allow the regular lattice tensor splines to efficiently apply to perform the functional data analysis (FDA) on the data $f_i$, $i=1,\dots,n$ defined on the $d$-dimensional domains.   

\section{Towards higher-dimensions}
\label{sec:towards}
In data analysis, if one wishes to apply 1D methods when dealing with images, it is necessary to convert two-dimensional (2D) data, such as pixel matrices, into one-dimensional (1D) sequences. This transformation is usually achieved by rearranging the pixels either column-wise or row-wise into a linear sequence. While this vectorization of image data simplifies the representation for processing and analysis, it also introduces significant limitations. Crucial spatial relationships and contextual nuances inherent in the original 2D layout are often lost in this flattened format. Consequently, only vertical or horizontal correlations are captured, and the rich, intricate nature of the image's two-dimensional character is substantially diminished, potentially leading to a loss of vital information for a comprehensive understanding of the data.
To minimize the negative effect of transforming from 2D to 1D, one can use more subtle domain or space transformations in an attempt to capture the local correlations between pixels. 
Two common transformation methods are often used for this purpose. One transforms the domain using the Hilbert curve, and the other transforms the value by using the gradient.

\subsection{Hilbert curve -- domain transform}
To address this limitation and better preserve local correlations, the Hilbert curve transformation is employed. The Hilbert curve is a continuous fractal space-filling curve that systematically passes over every point within a square grid and can be adjusted to any size with a 2-power magnitude. For a detailed exploration of Hilbert curves, a comprehensive reference can be found in \cite{bader2012space}. Hilbert curves are especially intriguing because of their capacity to group nearby pixels together.

In Figure~\ref{HCGradient}, in the right column, we see the difference between two image vectorizations: the one using stacking columns (middle) and the one using the Hilbert transform (right).   
We observe a less noisy character of the Hilbert curve data.



\subsection{Gradient images -- state transform}
As known from the success of convolutional methods in image analysis, one can efficiently capture some local dependence by using filters. 
Filtering an image captures local structures through transformation of the values of an image rather than its domain which was the case in the previous method. 
One of the most important filters is the gradient filter, which is equivalent to extracting the gradient image as discussed next.

The gradient images are generated by computing gradients representing local changes across the pixels of the original image. The gradient image effectively highlights regions where there are significant transitions or variations in pixel values, thus emphasizing edges, contours, and areas with notable curvature.
The gradient $G$ at the pixel $p=(i,j)$ of the image $I$ is defined as 
\begin{align*}
  G(p)
  &=
  \sqrt{\Delta^2_{x,p}+\Delta^2_{y,p}}\\
  &=\sqrt{
  \left(\frac{I(i+1,j)-I(i-1,j)}2\right)^2+
  \left(\frac{I(i,j+1)-I(i,j-1)}2\right)^2
  }.
\end{align*}
This operation combines the information from both horizontal and vertical changes in pixel values to determine the overall magnitude of change in intensity or color at each pixel location.

By analyzing gradient images, we can gain valuable insights into the spatial distribution of image features, enabling us to identify regions where the data exhibit significant curvature. Consequently, we obtain a two-dimensional gradient image denoted as $G$. 
This image can be viewed as a new functional data point that can in turn be transformed into a one-dimensional representation and treated jointly with the original image $I$.
The spatial information is now coded through the gradient which guides differently to strategically place knots when applying the DDK method.

\begin{figure}[t!]
  \centering
\raisebox{-0mm}{\includegraphics[width=0.32\textwidth]{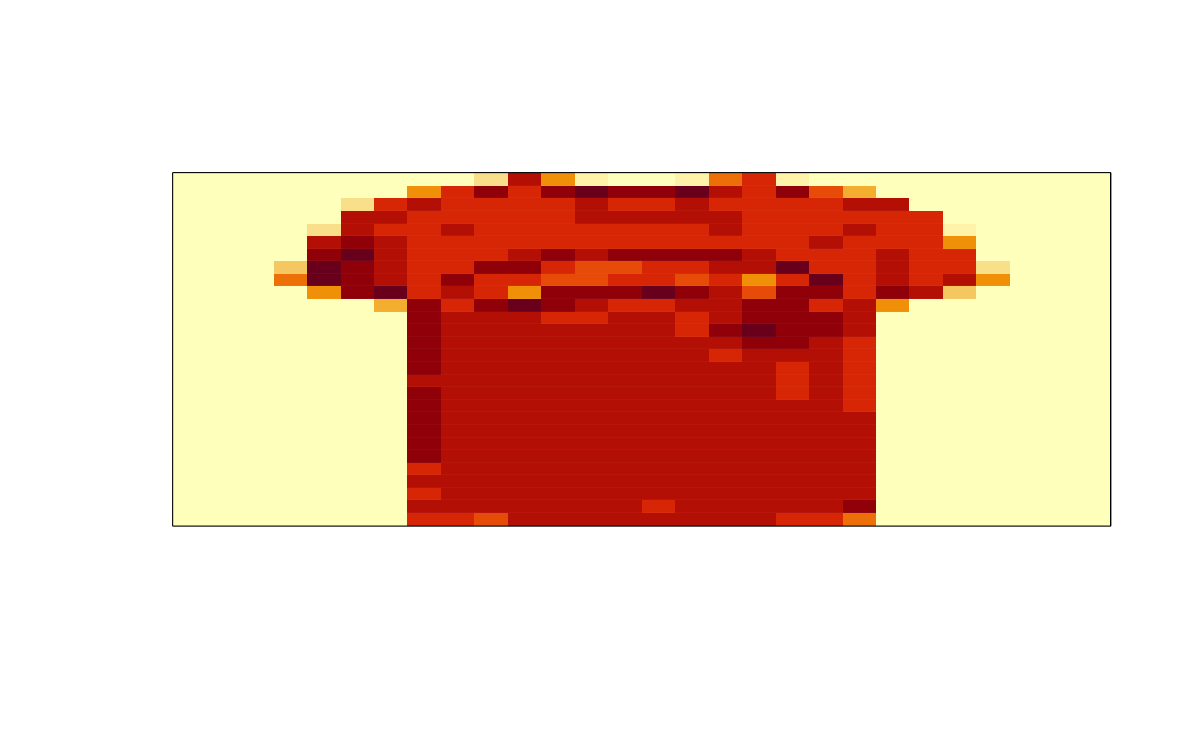}} \hspace{-2mm}
\includegraphics[width=0.33\textwidth]{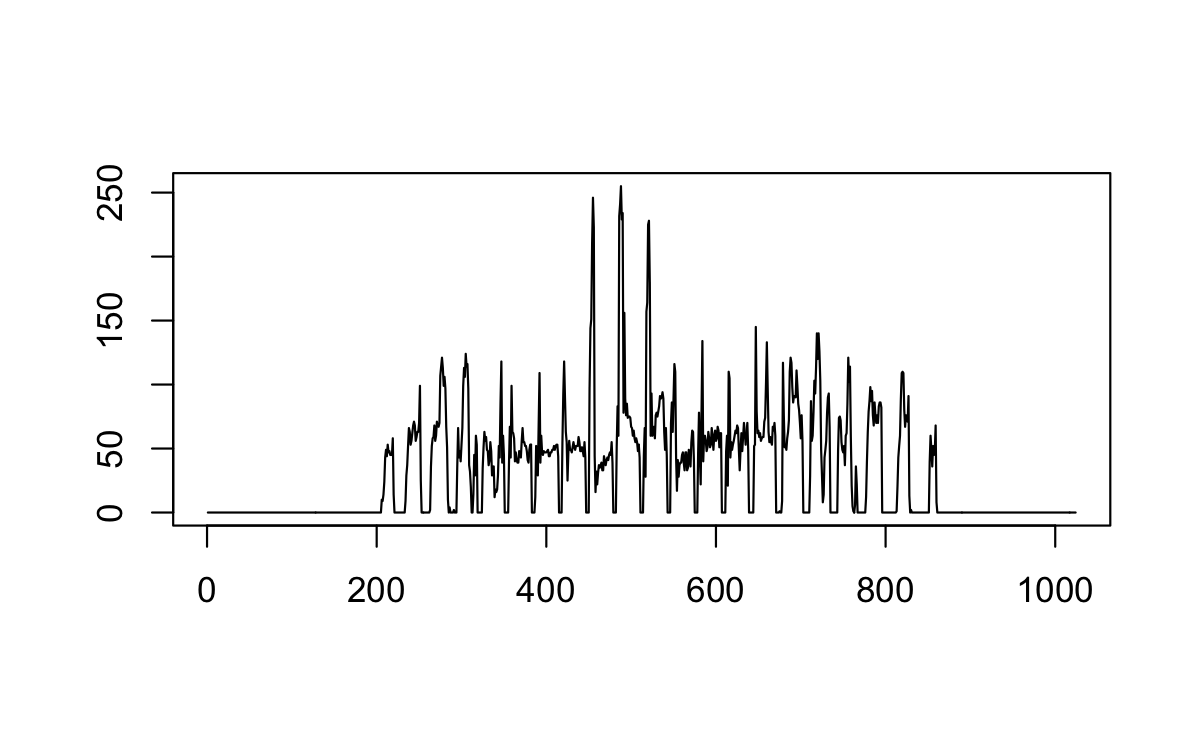}
\includegraphics[width=0.33\textwidth]{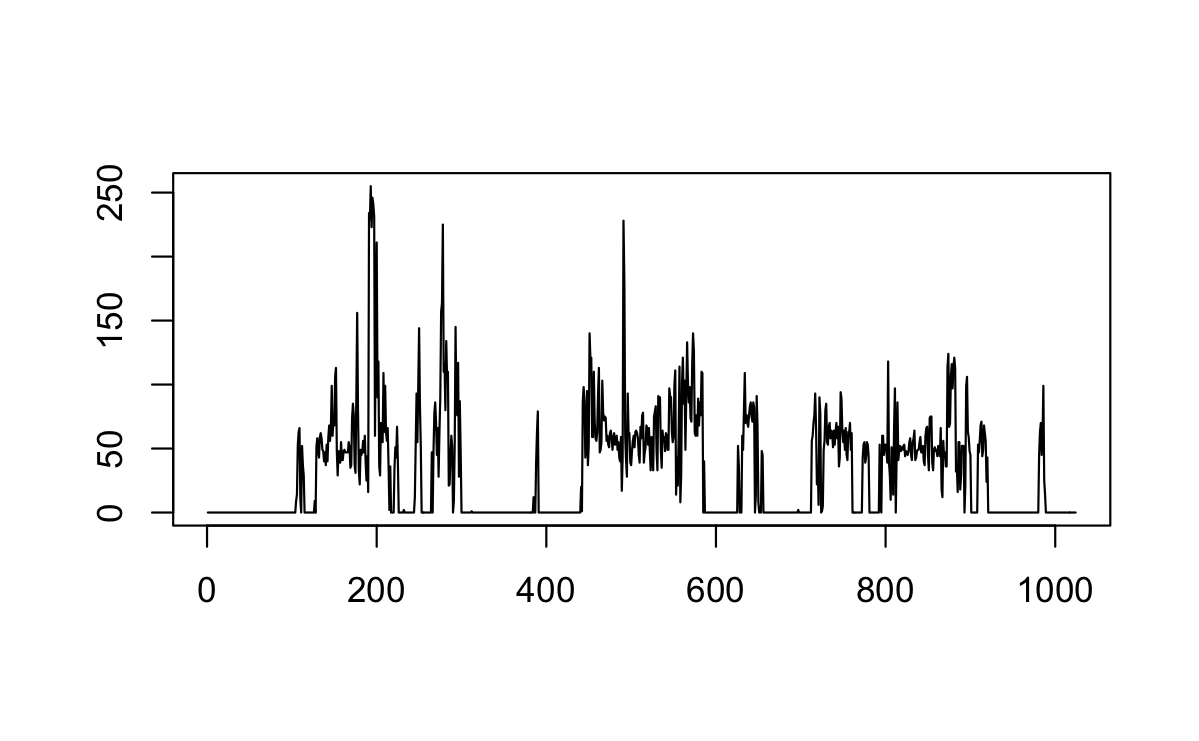}\\
\vspace{-0.5cm}
\raisebox{-0mm}{\includegraphics[width=0.32\textwidth]{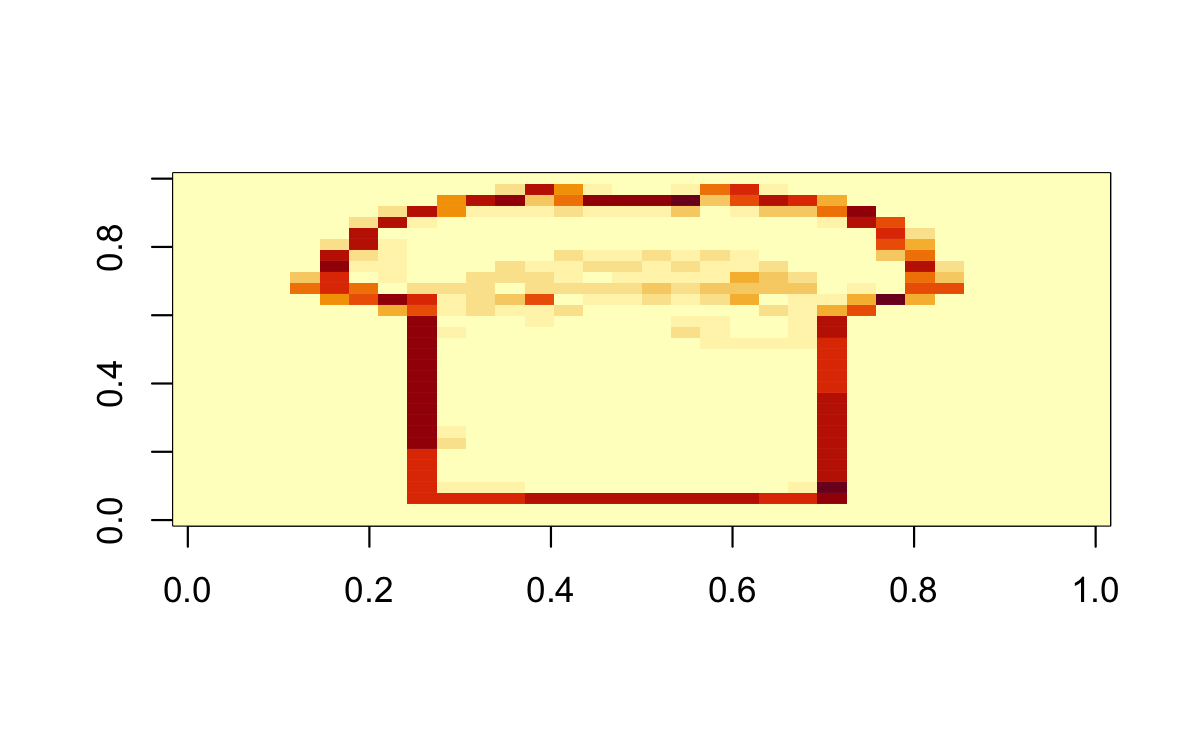} }\hspace{-2mm}
\includegraphics[width=0.33\textwidth]{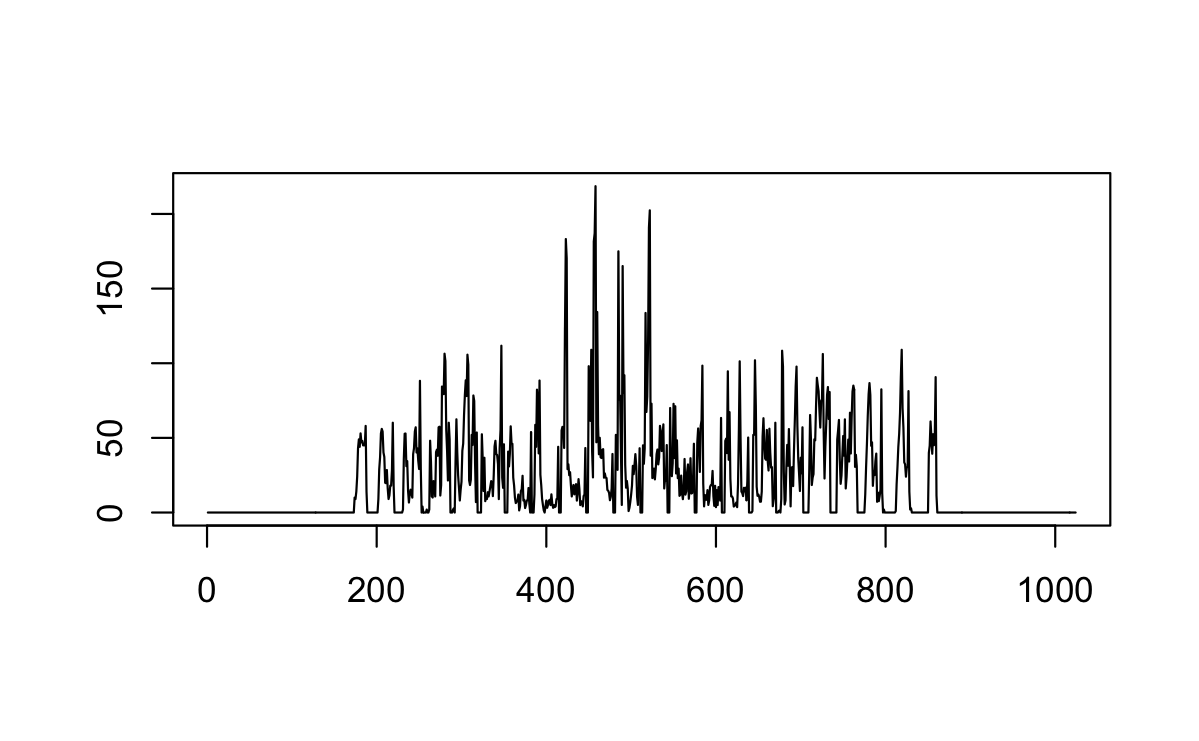}
\includegraphics[width=0.33\textwidth]{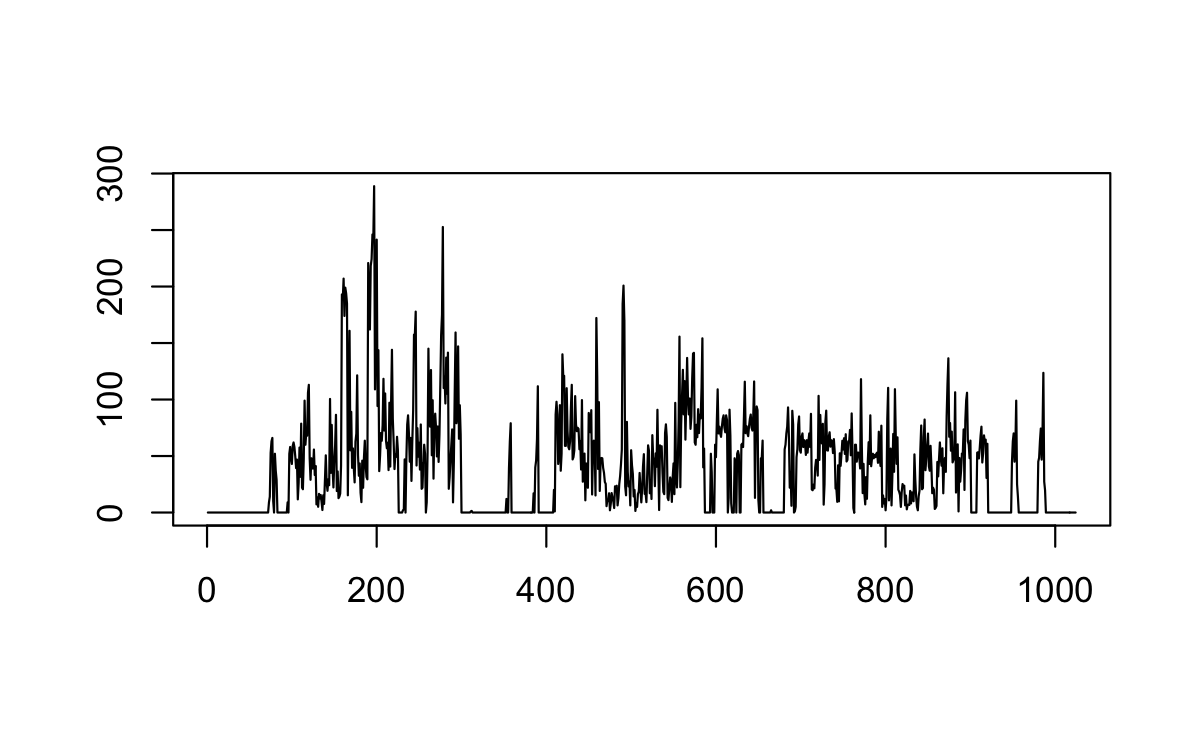}
  \caption{One- and two-dimensional representations of an example of MNIST data, a T-shirt and its gradient. {\it Left:} original images, 
  {\it Middle:} column-major order representation.
  {\it Right:} Hilbert curves representations of the T-shirt and its gradient.}
  \label{HCGradient}
  \end{figure}

In Figure~\ref{HCGradient}, we present two illustrative examples sourced from the MNIST dataset: an image of a shirt and its gradient image. The middle illustrations depict the vectorized representations of this item and its gradient, obtained by stacking the columns of the image matrix consecutively. In contrast, the left-hand-side depictions showcase vectors generated through the Hilbert space-filling curve. Notably, the Hilbert curve's ability to preserve local relationships is evident, making it a superior choice when compared to other transformation methods.

\begin{figure}
    \centering
    \includegraphics[width=0.5\textwidth]{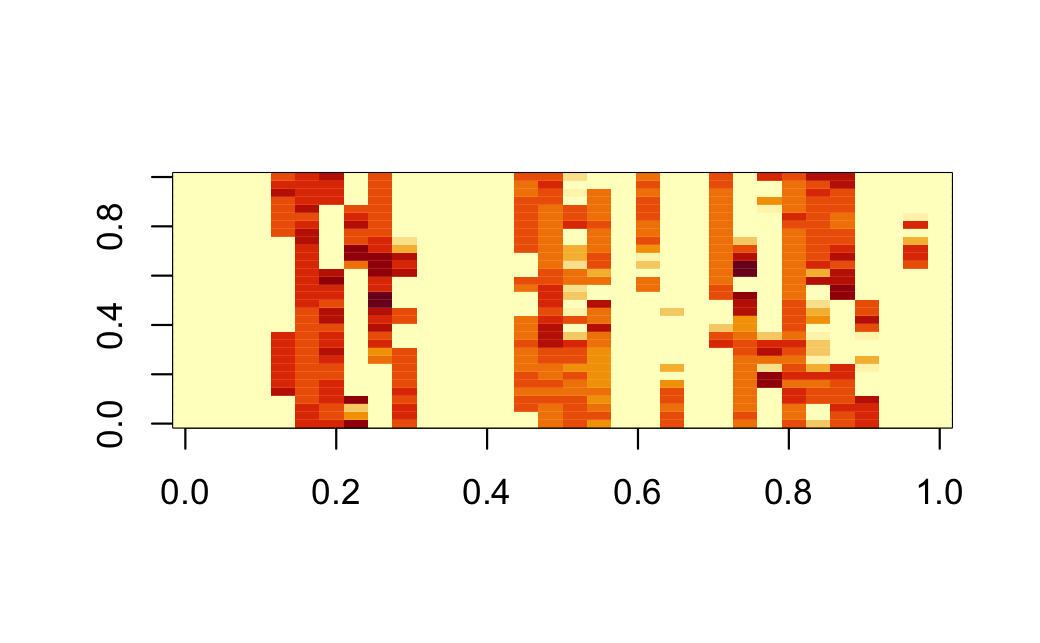}
    \caption{Caption}
    \label{fig:T-shirtHcImage}
\end{figure}

\subsection{FDA Classification workflow}
\label{sec:workflow}
Next, we examine the impact of employing these two methods in an example of classification 2D data analysis based on the functional principle component analysis.
The classification workflow has been originally presented in \cite{2023BasnaRP}, where also the effects of using the Hilbert transform have been quantified. We conduct a classification task on the Fashion MNIST dataset using gradient images
transformed into one dimension via the Hilbert curve. Unlike the approach in
\cite{2023BasnaRP}, which applies classification to the original images of the Fashion MNIST
dataset, our focus is on gradient images.

The workflow for this classification problem integrates a machine learning-based method
for data-driven knot (DDK) selection, as proposed in \cite{basna2022data}, and utilizes
the effective features of the Splines package for analyzing functional data,
as demonstrated in the \href{https://cran.r-project.org/web/packages/Splinets/index.html}
{\tt Splinets} package.

In the knot selection process, our aim is to achieve data-driven functional space
selection. This distinguishes our method from standard approaches that typically
rely on theoretical initial basis selections, such as Fourier basis, wavelets,
or splines with equidistant knots.

The entire workflow is summarized in the following steps (for more details on the workflow, see \cite{2023BasnaRP}. Moreover, the code for this workflow is available as supplementary material to the previously mentioned paper \cite{2023BasnaRP}.):

\begin{enumerate}
\item \textbf{Data preparation:} Split datasets into training, validation,
and testing parts, restructuring data representations as needed. Here,
the Hilbert curve algorithm is applied to gradient images for one-dimensional
data transformation.
\item \textbf{Data-driven knot selection:} Adopt the choice of spline spaces 
effectively. Apply the DDK algorithm to select class-specific knots in the
training dataset, determining both the number and placement of knots.
\item \textbf{Projecting training data into spline spaces:} After selection of the knot,
data are projected into the spline space defined by these knots, tailored to
each class of a training data point. Choose the order of spline spaces based
on requirements; in this case, order 2 for images and order 1 for gradient images.
\item \textbf{Classwise functional principal components analysis (FPCA):} 
Perform FPCA on training data within each class. This involves the standard
eigenvalue-eigenvector decomposition of positive definite matrices, aided by
the isometry between coefficients in the spline representation of a spline
and an Euclidean space of corresponding dimension.
\item \textbf{Determination of significant principal components:} Further reduce 
dimensions by selecting significant eigenvalues based on classification 
accuracy on the validation dataset. Choose the optimal number of eigenfunctions 
for best classification accuracy.
\item \textbf{Testing classification procedure:} Apply the utilized 
classification procedure discussed in \cite{2023BasnaRP} to the testing set, following the selection of all hyperparameters, including the number of eigenvectors. Classify each testing 
data point based on the closest projection to the principal component space 
of each class.
\item \textbf{Final evaluation and conclusions:} This final stage is tailored to the specific goals of the analysis and varies case by case.
\end{enumerate}

In the second step of our workflow, which involves data-driven knot selection, the algorithm has the potential to be extended to two dimensions for optimal knot placement. One feasible approach to generate the spline basis in two dimensions is through the tensor product of two sets of one-dimensional splines. This is accomplished by taking the coordinates of each selected knot.

However, a notable drawback of this method is the potential for generating an excessive number of basis functions in certain areas, which may not be necessary. This issue primarily arises due to the nature of the tensor product. For instance, if a small area in the horizontal direction contains a high density of knots, the tensor product will result in a disproportionately large number of basis functions spanning across this area along the entire vertical direction. This can lead to an uneven distribution of basis functions, potentially affecting the efficiency and accuracy of the spline representation in capturing the essential features of the data.

To mitigate this issue, it may be necessary to involve adaptive algorithms that take into account the density and distribution of selected knots, ensuring they effectively capture the features within the data. This approach aims for a more uniform and effective representation across both dimensions. Subsequently, we propose two different methods to address these challenges.

 \section{The state space transformation}
\label{sec:state} 
 The problem of irregularly spread `knots' occurs only for domains of a dimension higher than one. 
Our proposed methods account for `knots' distribution not directly in the construction of spline bases and can be applied in 1D as well. 
In this case, the irregular knots can be used directly in building splines thus one can benchmark the proposed methods against this direct irregular knot selection.

 The main component of both methods is a continuous approximation of the irregular `knot' distribution. 
 We consider `knots' $\boldsymbol \xi_j$, $j=1,\dots, N$ that are irregularly distributed. 
One can simply apply kernel-driven estimation-like methods known in non-parametric statistics by treating the `knots' as iid data drawn from a certain distribution. 
Since `knots' are not truly iid data, those methods should be viewed as ad hoc smoothing methods.
More suitably, one can use {\it Splinet}-package to project discretized data to the space of splines as presented in Theorem~\ref{th:project}. 
From now on, $g$ represents the density of the `knot' distribution.
 
Small values of $g$ indicate regions where the data are less informational. This motivates us to introduce the $g$-weighted innerproduct topology $\langle\cdot,\cdot\rangle_g$ in $L^2_g=\{h: \int h^2 g <\infty\}$  
$$
\langle h,k \rangle_g = \int h(x) k(x) g(x) dx = \int  h(x) \sqrt{g(x)} k(x)\sqrt{g(x)}~ dx = \langle h \sqrt{g},k \sqrt{g}\rangle. 
$$
We can transfer the functional data $X_1, X_2, \dots, X_n $ to 
\begin{equation}
\label{weighted_data}
    \Tilde{X_i}(x) = X_i(x) \sqrt{g(x)}.
\end{equation}
One can use $\Tilde{X_1}(x), \Tilde{X_2}(x), \dots, \Tilde{X_n}(x)$ instead of the original data and perform any functional analysis with one important modification. 
Namely, since the knot distribution is already accounted for, one does not need to use splines built on irregularly distributed knots. 
This is crucial in the 2D case since the equally spaced knots constitute a lattice on which tensor splines are well defined. 
\begin{remark}
The sparsity of `knots' in some regions may lead to zero (or nearly zero) values of $g$. This may lead to a subdomain of the original $[a,b]\times [c,d]$. This subdomain could constitute the support of the functional data.  To benefit from that one would like to have the possibility to account for a reduced support set. This feature is actually implemented in the current version of {\it Splinets}-package.
\end{remark}

 \section{The domain space transformation}
\label{sec:domain}
For the sake of simplicity of the exposition, we consider here only 2D and 1D cases. 
However, the approach is valid in any dimension. 
We also consider the `knot' distribution but instead of the density $g$, we utilize the distribution function $G(x,y)=\int_a^x\int_c^y g(u,v)~du~dv$. 
For that $g$ approximated in the previous method can be used. 
Alternatively, one can approach the problem directly by looking at the `empirical' equivalent of $G$:
$$
\hat{G}(x,y) = \frac{\#\{{\boldsymbol \xi}\cap [a,x]\times [c,y] \}}{N}.
$$
Then $\hat{G}$ is used to construct a domain mapping $\mathbf G$ from $[a,b]\times [c,d]$ to $[0,1]\times [0,1]$ such that if $\boldsymbol \xi$ is distributed according to the density $g(u,v)$, then $\mathbf G(\boldsymbol{\xi})$ is  distributed uniformly on the unit square.

To simplify the discussion, we turn to the 1D case.
Our functional data point, say $X$, defined on $[a,b]$, is transformed to  $\tilde X= X\circ G^{-1}$ on $[0,1]$.
These domain-transformed data are assumed to `live' in the Hilbert space $L^2$ of square integrable functions with the regular inner product $\langle \dot ,\dot \rangle$. Thus, for the elements $\tilde h, \tilde k$ in $L^2$, we have 
\begin{align*}
\langle \tilde h , \tilde k \rangle&=\int_0^1 h(G^{-1}(u)) k(G^{-1}(u))~ du= \int_a^b h(x) k(x) {g(x)} ~ dx
=\langle h, k\rangle_{g}
\end{align*}
We conclude that while the two methods transform data in different ways they are topologically equivalent, which we summarize in the following result. 
\begin{theorem}
\label{th:equiv}
Let $g$ be the density distribution on $[0,1]$, $G$ be the corresponding cdf. Moreover, let $h_g= \sqrt{g}h$ for $h\in L_g^2$ and $\tilde h_G=h \circ G^{-1}$, then 
$$
\langle h,k\rangle_g=\langle h_g, k_g\rangle=\langle \tilde h_G, \tilde k_G\rangle, \,\, h,k\in L_g^2.
$$
\end{theorem}
\begin{remark}
We observe that the method of knot selection implemented in DDK and utilized in \cite{BasnaR2021} exhibits the importance of the choice of knots $\xi_i$, $i=0,\dots,N$ for spline representation of the data. 
If we interpret the knots in the domain transform approach, we obtain $\nu_{i}={G}(\xi_{i})$, $i=0,\dots,N$ that are approximately uniformly distributed, since $G$ is chosen to fit the empirical distribution dictated by the knot selection. Thus one can choose $\nu_i$'s equally spaced instead. This is the fundamental fact that allows us to transfer the benefit of a non-uniform knot selection to the uniform one using the transform method.  Further, due to the topological equivalence presented in Theorem~\ref{th:equiv}, this benefit is also inherited by the state transform method. 

\end{remark}
\section{Implementations within the tensor spline spaces}
\label{sec:impl}
Despite this topological equivalency of the two Hilbert spaces, there is a fundamental difference in how these two approaches are implemented when it comes to the tensor splines approximation. 

The state space transformation method exploits the linear isomorphic embedding
\begin{align*}
\Gamma:(L_g^2,\langle \cdot, \cdot\rangle_g)
&\mapsto 
(L^2,\langle \cdot, \cdot\rangle),\\
\langle f, k \rangle_g&=\langle \Gamma f,\Gamma k \rangle,
\end{align*}
given by $\Gamma f=f\sqrt{g}$. 
This transformation is a bijection if $g>0$ almost everywhere and $\Gamma^{-1} f=f/\sqrt{g}$. 
It is assumed that the functional data $X_i$'s come from $L_g^2$ and the  `correct' topology is given by $\langle \cdot, \cdot\rangle_g$. Since the tensor spline spaces are developed within $(L^2,\langle \cdot, \cdot\rangle)$ in order to decompose data in the tensor splines, the data are first transformed to $\tilde X_i=\Gamma X_i$ and the tensor splines with regularly spaced knots on a lattice are used for the tensor spline base projections of $\tilde X_i$'s.

\begin{figure}[t!]
\centering
\includegraphics[width=0.49\textwidth]{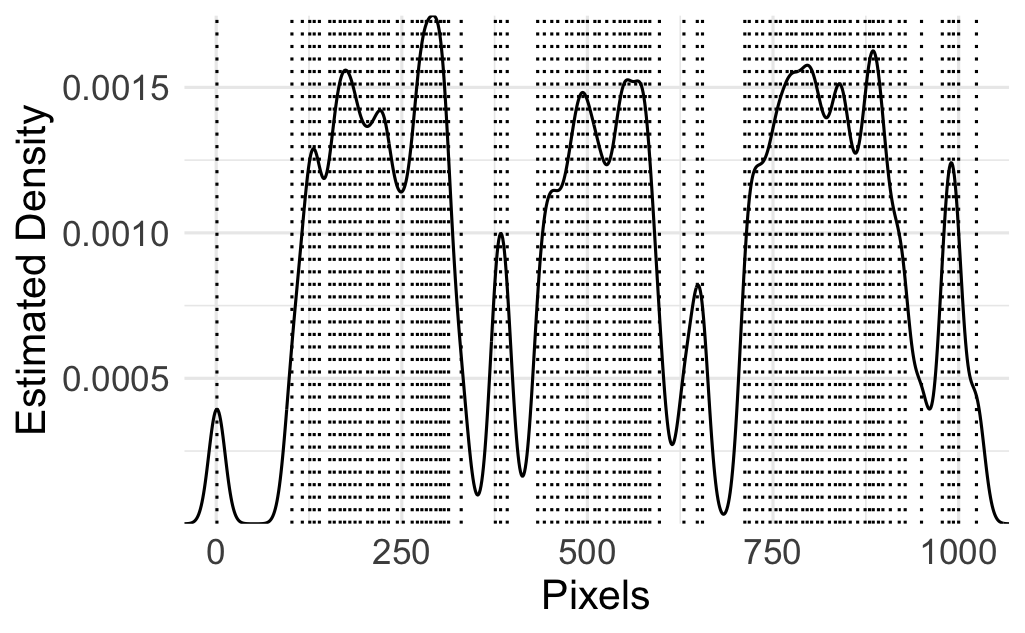}
\includegraphics[width=0.49\textwidth]{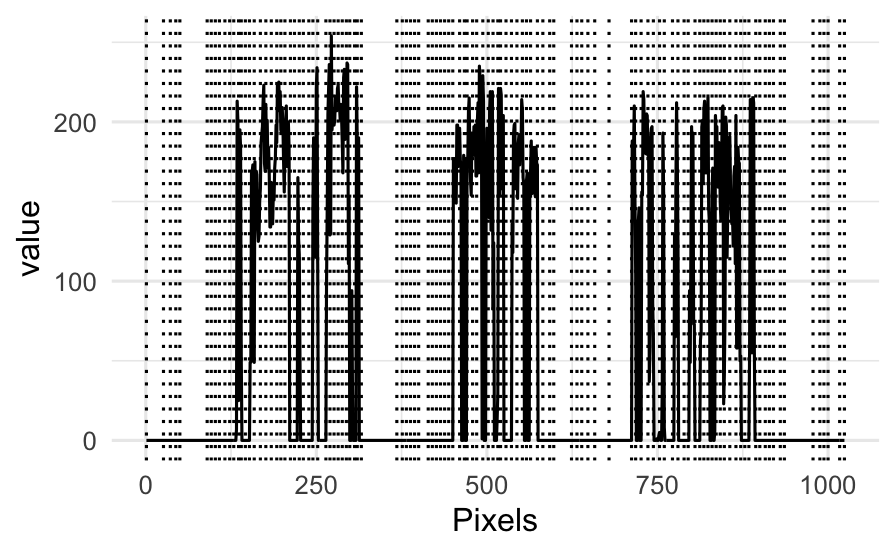}\\
\vspace{1cm}
\includegraphics[width=0.49\textwidth]{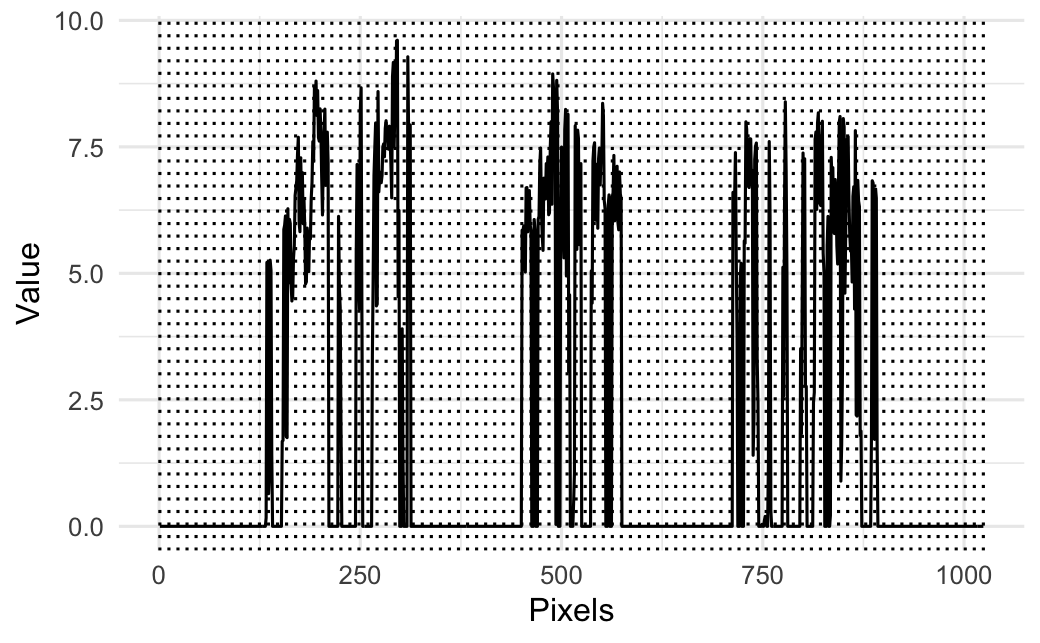}
  \caption{{\it Top Left}: Approximated density of knots for the data-driven approach selected with the DDK package for the Trouser class. {\it Top Right}: One-dimensional Hilbert curve representations of one sample from the Trouser class from the MNIST data. {\it Top Bottom}: One dimensional knots-density-weighted Hilbert curve representations of one sample from the Trouser class. 
   }
  \label{DE}
  \end{figure}

Since $g$ is interpreted as the intensity of importance map for the space on which it is defined, the knots on a regularly spaced lattice are placed irrespective of the importance given to their locations by $g$. 
In this sense, the dimensionality of the spline spaces (given by the number of their knots) is not affected by the method even if there may be many regions with importance close to zero.
We see in Figure~\ref{DE}{\it (Bottom)} that the equal placement of knots for splines for a trouser data point is the same irrespective of whether the curve vanishes or not. 

The role of tensor splines in the domain transformation method is different. Let us first explicitly formulate the embedding transformation. 
This time the linear mapping $\Psi f=f\circ G^{-1}$ defines isomorphic relation
\begin{align*}
\Psi:(L_g^2,\langle \cdot, \cdot\rangle_g)
&\mapsto 
(L^2,\langle \cdot, \cdot\rangle),\\
\langle f, k \rangle_g&=\langle \Psi f,\Psi k \rangle.
\end{align*}
Thus, similarly, as before the data $X_i$'s are residing in $L_g^2$ and they are transformed to $\check{X}_i=\Psi X_i$'s which resides in $L^2$ with the regular inner product. 
Also, both for $\tilde X_i$'s and $\check{X}_i$'s, the knots on an equally spaced lattice are used for the spline space approximation. 
However, these splines approximations do not lead to the same approximating functions as the following relations show. 
If $s_N$, $\tilde s_{\tilde N}$ is a spline spanned on equally space knots $\xi_i=i/N$, $i=0,\dots,N$, $\tilde \xi_i=i/\tilde N$, $i=0,\dots,\tilde N$, respectively, that approximates $\Psi f$, $\Gamma f$, respectively,  with the $\epsilon>0$-accuracy, so that
\begin{align*}
\epsilon &=\|\Psi f-s_N\|=\left(\int_0^1 \left( f(G^{-1}(u))-s_N(u)\right)^2~du\right)^{1/2}
\\
&=\left(\int_a^b \left( f(x)-s_N(G(x))\right)^2~dG(x)\right)^{1/2}
,\\
\epsilon &=\|\Gamma f-\tilde s_{\tilde N}\|=\left(\int_a^b \left( f(x)\sqrt{g(x)}-\tilde s_{\tilde N}(x)\right)^2~dx\right)^{1/2}
\\
&=\left(\int_a^b \left( f(x)-\tilde s_{\tilde N}(x)/\sqrt{g(x)}\right)^2~dG(x)\right)^{1/2}
.
\end{align*}
One can observe that both $s_N\circ G$ and $\tilde s_N/\sqrt{g}$ approximate $f$ in the same topology, with the same accuracy, and have the splines, $s_N$ and $\tilde s_{\tilde N}$, respectively, with equally spaced knots behind the approximations. 
However, they are fundamentally two different functions, that are, in general, not even splines. For example, there is no guarantee that $N=\tilde N$, i.e. the same number of knots will yield the same accuracy, thus one method may still have an advantage in the dimension reduction if it requires less knots to obtain a good approximation. 
We have commented above that the state transformation method may lead to an unnecessary too many knots in the data representation of sparse data so possibly the domain transformation method may have an advantage in this aspect. However, more studies are needed to confirm this claim. 

\section{Classification for functional data on a \texorpdfstring{$d$-}dimensional domain}
In the context of functional analysis, consider a $d$-dimensional functional space, denoted as $\mathcal{H}$, with its domain represented by $\mathcal{D} = \bigotimes_{j=1}^d [a_j,b_j]$. This space is equipped with an inner product $\langle x, y \rangle$ for any $x, y \in \mathcal{H}$. Our goal is to develop a comprehensive classification framework for functional data defined over this domain $\mathcal{D}$.

The classification methodology presented in \cite{2023BasnaRP} was initially formulated for one-dimensional settings but can be generalized to higher-dimensional domains. This section outlines the application and adaptation of this methodology to functional Hilbert spaces, $\mathcal{H}$, in which their elements (functions) are having domain of an arbitrary finite dimension.

Consider functional data in a $d$-dimensional space, encompassing $K$ distinct classes, denoted as $X_i(t)$ where $t \in \bigotimes_{j=1}^d [a_j,b_j]$. 
The fundamental novelty proposed here is that the class $k$ has its own  space $\mathcal H_k$ with the inner product $\langle\cdot,\cdot \rangle_{g_{k}}$ in which it is living.
However, the data can be transformed as described in Sections~\ref{sec:state} and~\ref{sec:domain} to the common space in which the regular topology of $L_2$ and equally spaced splines are considered.
This transformation is class-specific and depends on the `knot' intensities $g_k$, $k=1,\dots,K$.

As a preliminary step, the dataset is partitioned into training, validation, and testing sets. For each class within the training data set, we then proceed to construct suitable functional spline spaces, denoted as $\mathcal{S}_k$. Ideally, $\mathcal{S}_k$ is established as an orthogonal basis, with each element in this basis formulated as a tensor product of splinet bases, as discussed in Section \ref{sec:tensor}. These splinet bases are orthogonal with respect to the inner product defined in $\mathcal{H}$.

Then one projects the discretized training data in each class into the respective spaces; these functions from now on are referred to as functional data points. Subsequently, functional principal component analysis (FPCA) is performed within each class of the training set to build principal components within the class-specific Hilbert space $\mathcal H_k$, or, equivalently, as shown before, in the common space but then a functional data point needs to be transformed using the class-specific $g_k$, $k=1,\dots, K$ with $K$ denoting the number of classes.

This involves computing the means for each class, and the data, after centering around these means, are decomposed into eigenvalues and eigenfunctions. This results in spectral decomposition for each class separately, entailing the estimation of eigenvalues ${\lambda_{k1} > \lambda_{k2} > \dots > \lambda_{kn_k}}$ for the $k$th class and the corresponding set of eigenfunctions ${e_{k1}(t), e_{k2}(t), \dots, e_{kn_k}(t)}$, where $n_k$ represents the number of eigenfunctions per class $i$, and $k=1, 2, \dots, K$. To determine the optimal number of eigenvalues and eigenfunctions, a validation method is used.

The theoretical justification of this spectral decomposition is in alignment with the Karhunen-Lo\`eve's representation of functional data.   This representation asserts that functions within each class are independently sampled from the following model
\begin{align}
\label{eq:spectdec}
X_k(t) = \mu_k(t) + \sum_{i=1}^\infty \sqrt{\lambda_{ki}} Z_{ki} e_{ki}(t),, k=1, \dots, K,
\end{align}
where, $Z_{ki}$'s represent mean-zero, variance-one variables that are uncorrelated within a class and independent between classes.

The central principle of classification revolves around assessing how closely a data point $f(t)$ aligns with the projection onto the eigenspaces associated with a specific class. This involves the application of mathematical techniques that include feature extraction, dimensionality reduction, and the use of classification algorithms. These techniques are crucial for extracting meaningful insights from functional data points and efficiently categorizing them.

Let us give more details of this classification procedure. 
 In the training step, the 
mean function for each of the $K$ classes, utilizing our training data sets. For a given class, the mean function of all elements in the training set belonging to the $i$th class is denoted by $\hat{\mu}_i$. This function is considered an estimated value of the true mean $\mu_i$. It is obtained by projecting the averaged data corresponding to that class into the respective functional spline space, $\mathcal{S}_i$.

For each original data point $x_l$ in the test set, we evaluate its transformations using $g_k$, $k=1,\dots, K$ and then representations $f_{lk}$, where $k=1,\dots, K$, in the respective functional spline space, $\mathcal{S}_i$. Our objective is to assess how closely each transformed data point $f_{lk}$ aligns with its respective class by projecting $f_{lk}-\hat{\mu}_k$ onto the eigenspaces (the spaces spanned by eigenfunctions) corresponding to each class.
This process results in $K$ distinctive projections, which we denote as $\hat{f}_{l1},\hat{f}_{l2}, \dots, \hat{f}_{lK}$. More explicitly, 
\begin{equation}
\label{eq:eigennu}
 \hat{f}_{lk}(t) =\hat \mu_k (t)+ \sum_{i=1}^{n_k} \langle f_{lk
 k}(t)-\hat\mu_k(t),\hat e_{ki}(t) \rangle \hat e_{ki}(t),\, k=1,\dots, K.   
\end{equation}
Here, $\langle\cdot \rangle$ represents the inner product within the functional spaces, and $\hat e_{ki}$ are estimates of the eigenfunctions $e_{ki}$ obtained during the training phase, as previously discussed.

Formally, we obtain the following spectral decompositions
$$
f_{lk}(t)=\hat f_{lk}(t) +\hat\varepsilon_{lk}(t),
$$
where $\hat\varepsilon_{lk}$ represents the residual of the projection.

If the functional point $x_l$ belongs to the $k$th class, and $\hat \mu_k$ and $\hat e_{ki}$ are approximately equal to the true values, we can infer from \eqref{eq:spectdec} that
\begin{align*}
\hat\varepsilon_{lk}(t)
\approx 
\sum_{i={n_k}+1}^\infty \sqrt{\lambda_{ki}} Z_{ki}  e_{ki}(t),
\end{align*}
which should be relatively small when measured in the squared norm $\|\cdot\|$ of the functional spaces. In fact, we can approximate this as
$$
\|\hat\varepsilon_{lk}\|^2\approx
\left \|
\sum_{i={n_k}+1}^\infty \sqrt{\lambda_{ki}} Z_{ki} e_{ki}(t)
\right \|^2
=
\sum_{i=n_k+1}^\infty \lambda_{ki} Z_{ki}^2 .
$$

Consequently,
$$
\mathbb E (\|\hat\varepsilon_{lk}\|^2) \approx \mathbb E\left(\left \|
\sum_{i={n_k}+1}^\infty \sqrt{\lambda_{ki}} Z_{ki} e_{ki}(t)
\right \|^2 \right)
 \approx
 \sum_{i=n_k+1}^\infty \lambda_{ki}.
$$
This approximation can be made small as long as the selection of $n_k$ targets values such that
\begin{align*}
 \sum_{i=1}^\infty \lambda_{ki} &\approx   \sum_{i=1}^{n_k} \lambda_{ki}.
\end{align*}

Conversely, when $x_l$ does not belong to the $k$th class, we expect the residual $\hat\varepsilon_{lk}$ to be relatively large, especially when the classes are well-distinguished by projections onto the eigenspaces associated with their largest eigenvectors.

This observation justifies the following classification rule
\begin{equation}
    \label{eq:class}
    I(x_l)=\mathop{\rm arg\, min}_{k=1,\dots,K}\|\hat \varepsilon_{lk}\|=\mathop{\rm arg\, min}_{k=1,\dots,K}\|x_l -\hat f_{lk}\|,
\end{equation}
Here, $I(x_l)$ denotes the chosen class, and $x_l$ is treated as a piecewise constant function.

To further refine the classification outcome, we can calculate squared normalized distances, expressed as
\begin{equation}\label{eq:weights}
\left(w_1^l,\dots, w_K^l\right)=\frac{\left(\|x_l -\hat f_{l1}\|^2,\dots, \|x_l-\hat f_{lK}\|^2\right)}{\sum_{k=1}^K \|x_l-\hat f_{lk}\|^2}.
\end{equation}

These weights, denoted as $\left(w_1^l,\dots, w_K^l\right)$, provide a measure of the similarity between the test data point $x_l$ and each class's projection. They facilitate a more refined classification decision by taking into account the relative distances between $x_l$ and the projections onto each class's eigenspace.

\begin{remark}
The above presentation assumes some generic topology of the Hilbert space in which the functional data are represented. As we have discussed before, in our approaches not always the topology of the standard inner product is used. Thus in the above presentation the orthogonality and the norm $\|\cdot\|$ are meant to be in the proper topology.
For example, in the context of state space transformation, we consider a $g$-weighted topology, which is equipped with a weighted inner product. However, it is important to note that a weighted inner product in this space is equivalent to scaling the data by $\sqrt{g}$. Consequently, this allows us to leverage the more familiar $ L^2 $ space for our analysis. In essence, by scaling the data appropriately, we can transform the problem into one that is more manageable within the $ L^2 $  space, while still preserving the integrity of the original weighted topology. Similar comments apply to the domain transform approach. 
\end{remark}

\section{Illustration of classification principle in spline space projections}
In this section, we illustrate the classification workflow in Subsection~\ref{sec:workflow}, focusing on the transformed data as discussed in the state space transformation Section~\ref{sec:state}. The primary objective is to demonstrate the efficacy of the proposed state-space transformation as a response to the challenges outlined at the end of Section~\ref{sec:workflow}. 
Since the workflow undergoes changes in the first four steps when using transformed images instead of the original images, and remains the same thereafter, our focus will be solely on those steps where changes occur while leaving a comprehensive application of the complete classification process to future studies. For this illustration, we will utilize the Fashion MNIST dataset. The workflow will be based on the images transformed to 1D functional data using the Hilber curve method and we will use the techniques discussed in \ref{sec:towards}:

\begin{enumerate}
    \item \textbf{Data preparation}: Since the aim of this section is to provide an illustration, we will work with the workflow in 1D, employing techniques discussed in \ref{sec:towards}, namely the use of Hilbert curves. We have utilized the Hilbert curve to transform the 2D image data of the Fashion MNIST dataset into a one-dimensional format, thereby facilitating functional data analysis.
    

    \item \textbf{Data-driven knot selection}: The DDK algorithm was utilized to identify class-specific knots in the dataset. This involved determining the optimal number and placement of knots to capture class-specific variability. The DDK algorithm incorporates a built-in stopping criterion based on the relative error reduction. For each class, the algorithm monitors the decrease in average mean square error against Gaussian noise fitting. Knot addition ceases when the relative error aligns with the reduction achieved in noise fitting. A universal reference noise curve, computed via Monte Carlo simulations, serves as an independent benchmark. For more details on the stopping algorithm see \cite{basna2022data}
    and \cite{2023BasnaRP}.
    \item \textbf{State space transformation}: A state space transformation is applied after selecting knots for each class. This approach involves weighting the data in each class according to the corresponding estimated knot density, thereby allowing us to work within the familiar $L^2$ space. The transformed data, as detailed in \eqref{weighted_data}, are then utilized for further analysis.
    \item \textbf{Spline space projection}: In this step, we benefit from the state space transformation. After transforming the data using the density of the knots selected for each class, we can project all the data into a single spline space, which is built on equi-spaced knots. In contrast to the workflow in \cite{2023BasnaRP}, where the data in each class are projected onto different spaces built on the corresponding selected knots, our approach involves projecting all the data into a single spline space.

    \item \textbf{Functional principal components analysis (FPCA)}: 
    The projections made in the previous step establish an isomorphism between the Euclidean vectors made of coefficients of the projection and the space of splines into which the original data have been projected. Per-class FPCA was conducted, by first computing the covariance matrix of the isometric vectors of coefficients corresponding to the splinet projection. Next, a spectral decomposition of the covariance matrix into eigenvalues and eigenvectors is performed. As a results, we obtaine the eigenvectors correspond to eigenfunctions.
\end{enumerate}

\begin{figure}[t!]
\centering
\includegraphics[width=0.52\textwidth]{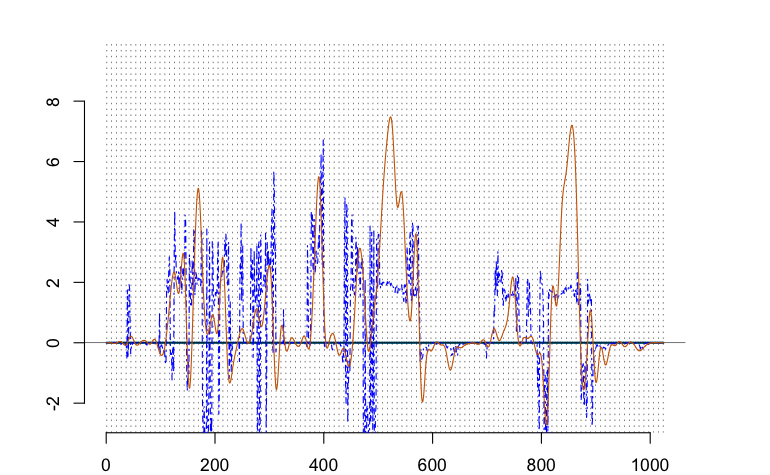}\hspace{-10mm}
\includegraphics[width=0.52\textwidth]{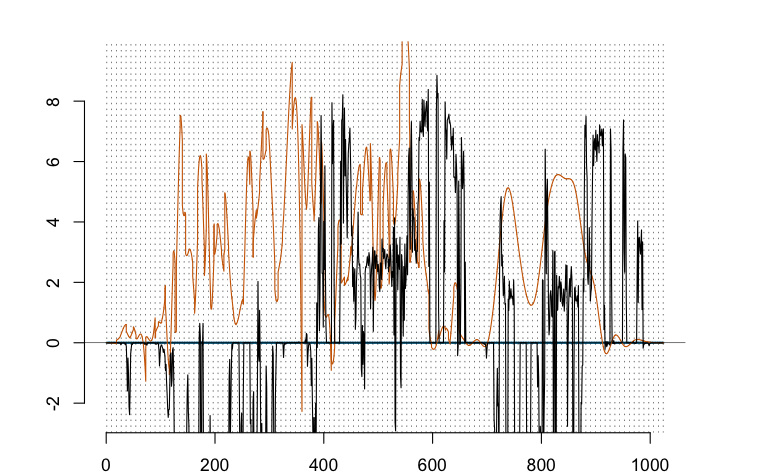}\hspace{-10mm}
  \caption{The spectral decomposition of the data. {\it Left}: Approximations of the centered scaled functional `T-shirt' data point {\it (blue-dashed line)}: by the first 20 eigenfunctions {\it (orange-solid line)}. {\it Right}: Projection of a `Boot' data point to the `T-Shirt' spectrum with 20 eigenfunctions after centering around the `T-Shirt' class mean, the discrete`Boot' data point {\it (black-rough line)} vs. the projection {\it (orange-solid line)}. 
   }
  \label{EigenFunctionApproximation}
  \end{figure}

Figure~\ref{EigenFunctionApproximation} illustrates the projection of two distinct data points — a `T-Shirt' {\it (left)} and a `Boot' {\it (right)} — into the eigenspace of the `T-Shirt' class. This visualization is the result of the procedural steps outlined above.
The projection of the `T-Shirt' image within its class-specific sub-eigenspace, spanned by the first 20 eigenfunctions, shows proximity to the original one-dimensional `T-shirt' scaled data. This indicates an effective capture of the class’s intrinsic variability through the state space transformation methodology.
In contrast, the projection of the `Boot' image exhibits significant deviation when placed in the `T-Shirt' sub-eigenspace, spanned by the first 20 eigenfunctions. This highlights the distinctiveness of each class’s eigenspace and underscores the effectiveness of our methodology in differentiating between classes. Overall, Figure~\ref{EigenFunctionApproximation}, exemplifies the potential effectiveness of our data-driven functional data analysis workflow in higher dimensional data classification. 

In summary, this example of an analytical framework, which includes knot selection, state space transformation, and FPCA, offers a promising foundation for effectively classifying high-dimensional image data within the scope of functional data analysis.

\bibliographystyle{unsrtnat}
\bibliography{RJreferences}